\documentclass[aps,prl,twocolumn,superscriptaddress,floatfix]{revtex4-1}
\usepackage{graphicx}
\usepackage{dcolumn}
\usepackage{bm}
\usepackage{color}
\usepackage{amsmath}
\usepackage{amssymb}
\usepackage{xcolor}
\usepackage{hyperref}
\usepackage{titlesec,setspace}
\usepackage[normalem]{ulem}
\usepackage{easyReview}

\usepackage{mathdots}

\newcommand{\bq}{{\bf q}}

\allowdisplaybreaks

\begin{document}
\title{Magnetic and charge susceptibilities in the half-filled triangular lattice Hubbard model}
\author{Shaozhi Li}
\affiliation{Department of Physics, University of Michigan,  Ann Arbor, Michigan 48109, USA}
\author{Emanuel Gull}
\affiliation{Department of Physics, University of Michigan,  Ann Arbor, Michigan 48109, USA}
\affiliation{Center for Computational Quantum Physics, The Flatiron Institute, New York, New York, 10010, USA}
\date{\today}

\begin{abstract}
We study magnetic and charge susceptibilities in the half-filled two-dimensional triangular Hubbard model within the dual fermion approximation in the metallic, Mott insulating, and crossover regions of parameter space.
In the \textcolor{black}{insulating state}, we find strong spin fluctuations at the K point at low energy corresponding to the \textcolor{black}{120$^{\circ}$} antiferromagnetic order. These spin fluctuations persist into the metallic phase and move to higher energy.
We also present data for simulated neutron spectroscopy and \textcolor{black}{spin-lattice} relaxation times,
and perform direct comparisons to inelastic neutron spectroscopy experiments on the triangular material Ba$_8$CoNb$_6$O$_{24}$ and to the relaxation times on $\kappa$-(ET)$_2$Cu$_2$(CN)$_3$. Finally, we present charge susceptibilities in different areas of parameter space, which should correspond to momentum-resolved electron-loss spectroscopy measurements on triangular compounds.
\end{abstract}


\maketitle
{\it Introduction}. Experimental evidence on several organic materials, including $\kappa$-(BEDT-TTF)$_2$Cu$_2$(CN)$_3$ \cite{Shimizu03,Kurosaki05}, EtMe$_3$Sb[Pd(dmit)$_2$]$_2$ \cite{Itou07,Yamashita10,Yamashita11}, and $\kappa$-H$_3$(Cat-EDT-TTF)$_2$ \cite{Isono14}, suggests that these compounds are close to a two-dimensional triangular structure and exhibit interesting electron correlation behavior including, potentially, a quantum spin liquid phase \cite{ZhouRMP2017} in the ground state \cite{Shirakawa17}.
These compounds, as well as the low energy physics of the fully isotropic triangular material Ba$_8$CoNb$_6$O$_{24}$ \cite{Rawl17}, may be described by a half-filled \textcolor{black}{single orbital} Hubbard model on a triangular two-dimensional lattice, with an on-site Coulomb interaction strength comparable to or larger than the bandwidth \cite{pustogow2018}.

Because of the subtle competition of metallic, ordered, and spin liquid phases in the ground state, this model has been studied extensively with a wide range of \textcolor{black}{numerical tools}, including exact diagonalization (ED) \cite{Koretsune07,Clay08,Kokalj13}, density matrix renormalization group theory (DMRG) \cite{Shirakawa17}, variational Monte Carlo (VMC) \cite{Watanabe06,Watanabe08,Tocchio14,Tocchio13}, variational cluster approximation \cite{Sahebsara08,Yamada14,Misumi17}, strong coupling expansions \cite{Yang10}, path integral renormalization group techniques \cite{Morita02}, and cluster dynamical mean field theory (DMFT) in the cellular \cite{Kyung06,Ohashi08,Liebsch09,Galanakis09,Sato12} and dynamical cluster \cite{Lee08,Dang15} variants. 
The focus in most of these studies has been on the precise location of the phase boundaries, ordering (or the absence thereof), and on the nature of these phases.

Experimentally, much of our knowledge about correlated triangular systems is obtained from single- and two-particle scattering experiments such as photoemission \citep{GePRB2014}, Raman spectroscopy \citep{LemmensPRL2006}, nuclear magnetic resonance (NMR) \citep{Kurosaki05,shimizuPRL2016}, or inelastic neutron scattering \citep{Rawl17,SayaNatureComm2017}. 
To understand these experimental results, it is necessary to calculate the corresponding response functions as a function of energy and momentum. For neutron spectroscopy and angular-resolved photoemission spectroscopy, in particular, both fine momentum and energy resolutions are desired. 
Such results are difficult to obtain, as computational methods formulated on finite lattices (such as ED, DMRG, and  cluster DMFT) provide limited momentum resolution.
In addition, quantum Monte Carlo approaches are impeded by a sign problem in frustrated systems \citep{IglovikovPRB2015}. \textcolor{black}{Results for these quantities are therefore often obtained from fits to quantum spin models, which are only justified in the large Coulomb interaction limit.}

In this paper, we provide results for the momentum and energy dependence of the spin and charge spectra of the half-filled triangular lattice Hubbard model. We use the dual fermion (DF) approximation, which is a diagrammatic extension of the DMFT and recovers continuous momentum dependence without suffering from a sign problem. We perform simulations from weak to strong interactions and systematically study spin and charge spectra in different areas of parameter space. We then relate our results back to experiments on triangular lattice compounds and calculate nuclear magnetic resonance relaxation times.

\begin{figure*}[t]
\center{\includegraphics[width=\textwidth]{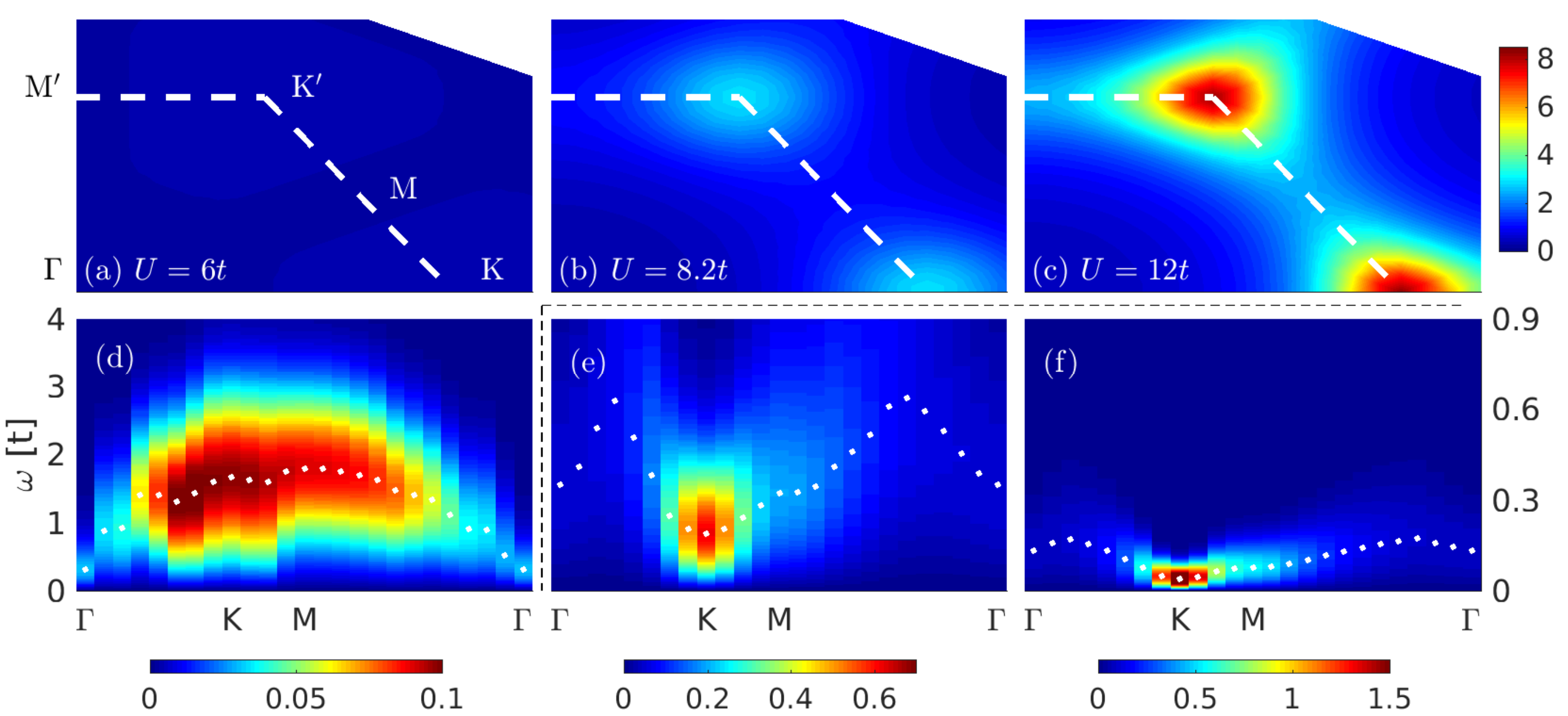}}
\caption{Panels (a), (b), and (c): Static magnetic susceptibility in the momentum space, $\chi_m({\bf q},\mathrm{i}\nu_0)$, for $U=6t$, $8.2t$, and $12t$ at $T=t/6$. The white dashed line delineates the Brillouin zone boundary. Panels (d), (e), and (f): Imaginary part of the dynamic magnetic susceptibility $\mathrm{Im}\chi_m({\bf q},\omega)$ along the high symmetry direction for the same values of $U$. White dots: peak position of the spectra $\omega_m({\bf q})$.}
\label{Fig:spin} 
\end{figure*}

{\it Model}. The Hubbard model is defined as
\begin{eqnarray}
H=-t\sum_{\langle ij\rangle,\sigma} \left( c_{i\sigma}^{\dagger}c_{j\sigma}^{\phantom\dagger} + h.c. \right) + U\sum_{i} \hat{n}_{i\uparrow}\hat{n}_{i\downarrow}.
\end{eqnarray}
$\langle \cdots \rangle$ denotes a summation over nearest neighbors; $c_{i\sigma}^{\dagger}$ ($c_{i\sigma}^{\phantom\dagger}$) creates (annihilates) an electron with spin $\sigma$ on site $i$; $\hat{n}_{i\sigma}=c_{i\sigma}^{\dagger}c_{i\sigma}^{\phantom\dagger}$ is the particle number operator; $U$ is the on-site Coulomb interaction; and $t$ is the nearest-neighbor hopping integral. We set $t=1$ throughout this paper and restrict ourselves to half filling.

{\it Method}. We study the model in the ladder dual fermion approximation \cite{Rubtsov08,Brener08,Li08,Hafermann12,Antipov14,Rohringer18} using the open source code of Ref.~\cite{ANTIPOV201543}.
The DF method is a diagrammatic extension of the DMFT \cite{RevModPhys.68.13} which treats all local correlations in a non-perturbative manner and perturbatively adds nonlocal charge and spin correlations \cite{Rohringer18,ANTIPOV201543}. DMFT calculations are performed with the continuous time auxiliary field quantum Monte Carlo method \cite{RevModPhys.83.349, Gull_2008} with submatrix updates \cite{PhysRevB.83.075122}.
DF is accurate at high temperature \cite{PhysRevX.5.041041} but uncontrolled in practice in the sense that adding systematic corrections, while possible in theory \cite{PhysRevB.96.235127,PhysRevB.94.035102}, is not feasible for the parameters studied here.
A detailed assessment of the approximation errors of the susceptibility and the single-particle properties on the square lattice \cite{LeBlanc19} showed that while doping- and interaction dependent scaling effects were present, the overall momentum dependence was accurate.

DF calculations are performed on a momentum space grid --- here we choose a square $24\times 24$ cluster, resulting in 288 points in the triangular lattice Brillouin zone. Both the single-particle Green's function and two-particle susceptibilities are defined on that grid. To examine the spectral properties, we use the ALPS implementation \cite{LEVY2017149,GAENKO2017235} of the maximum-entropy method \cite{JARRELL1996133} to perform the analytic continuation of Matsubara data to the real frequency space.

\begin{figure}[t]
\center{\includegraphics[width=\columnwidth]{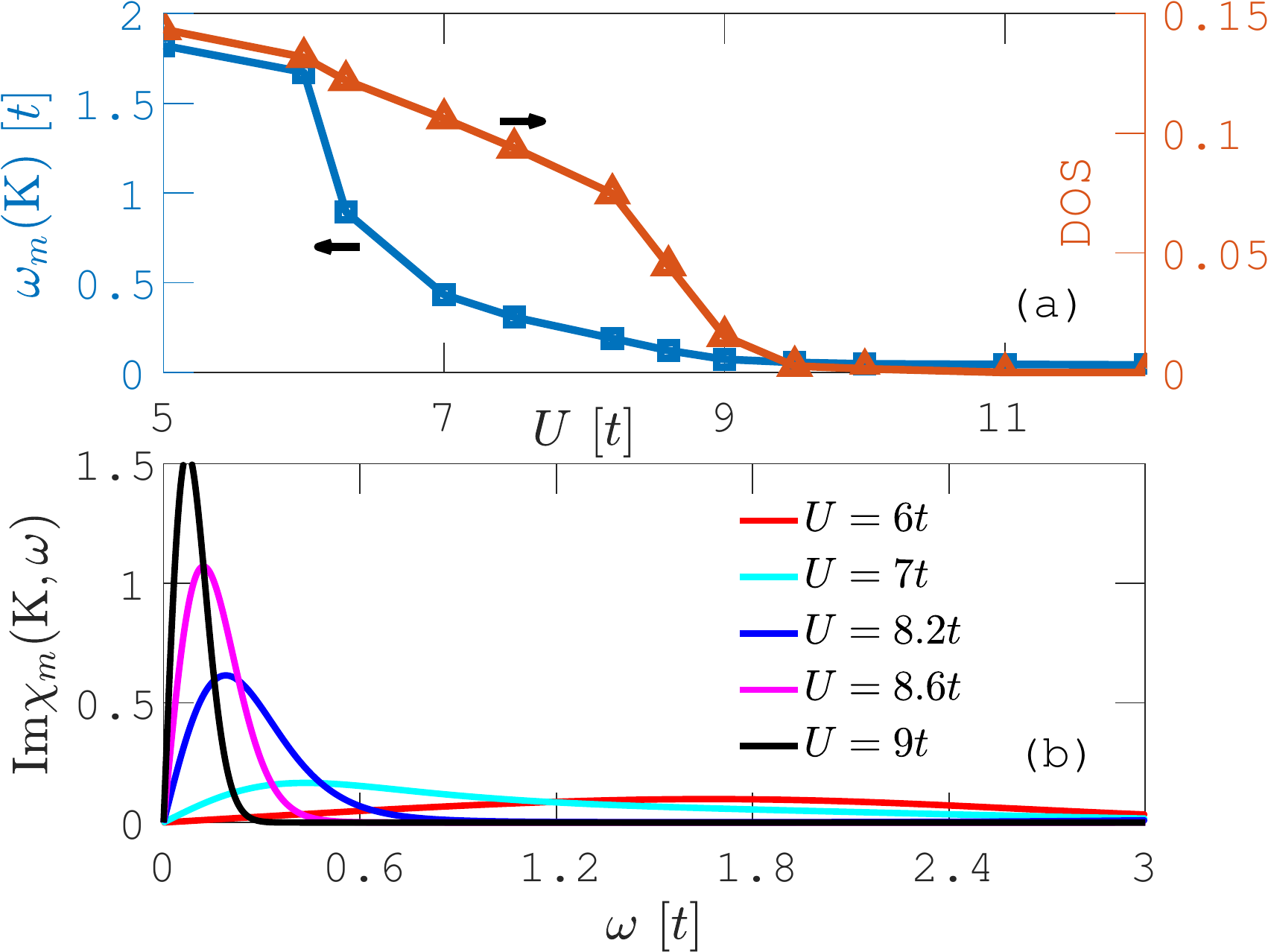}}
\caption{Panel (a): The spin excitation energy $\omega_m(\mathrm{K})$ at the K point and the \textcolor{black}{density of states} (DOS) at $\omega=0$ as a function of $U$. Panel (b): Imaginary part of dynamical magnetic susceptibility $\mathrm{Im}\chi_m({\bf q},\omega)$ at ${\bf q}=$K as a function of $\omega$ for various different $U$.}
\label{Fig:trace_k}
\end{figure}

{\it Results}. The half-filled Hubbard model exhibits a metal-insulator phase transition on the triangular lattice at low temperature, which has been widely studied by DF \cite{Li14,Laubach15,Lee08} and other methods \cite{Misumi17,Shirakawa17}.
In this work, we set \textcolor{black}{$T=t/6$}, which is above the critical temperature of this transition, such that the system exhibits metallic behavior for $U\le 8t$, crossover behavior for $8t<U<9.5t$, and insulating behavior for $U\ge9.5t$ \citep{Supplement}. 
To illustrate these three behaviors, we choose one point in each region: $U=6t$ in the metallic phase, $U=8.2t$ in the crossover region, and $U=12t$ in the insulating phase.

We first focus on the magnetic properties. Figure~\ref{Fig:spin} shows the momentum-resolved static and dynamical magnetic susceptibility for $U=6t$, $U=8.2t$, and $U=12t$. The static magnetic susceptibilities  $\chi_m(\bq,\mathrm{i}\nu_0=0)$ \cite{Supplement} for these three values of $U$ are plotted in \textcolor{black}{panels} (a), (b), and (c), respectively. The white dashed line represents the boundary of the Brillouin zone. Static spin correlations are enhanced as $U$ increases. At $U=12t$, the static spin correlations show a clear peak at the K point, which is much stronger than that in the metallic ($U=6t$) and crossover ($U=8.2t$) regions. The strong peak at $U=12t$ indicates the formation of \textcolor{black}{120$^{\circ}$} antiferromagnetic (AFM) spin fluctuations \cite{Shirakawa17}, which will magnetically order at low temperature \citep{Shirakawa17,Yoshioka09,Liebsch09}.

\begin{figure}[t]
\center{\includegraphics[width=\columnwidth]{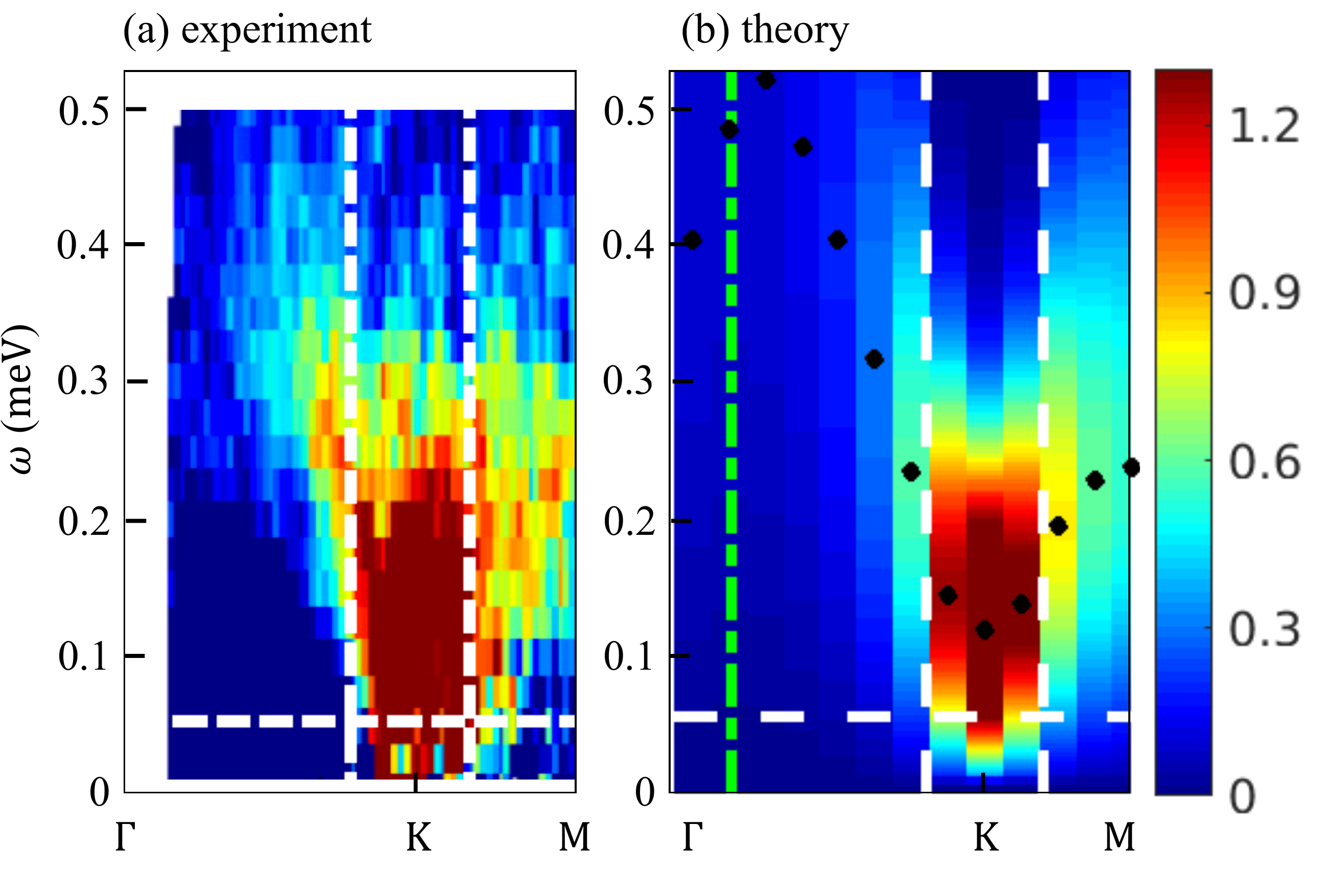}}
\caption{Panel (a): Spin susceptibility $\mathrm{Im}\chi_m({\bf q},\omega)$ obtained from inelastic neutron scattering on $\mathrm{Ba}_8\mathrm{CoNb}_6\mathrm{O}_{24}$ by \textcite{Rawl17} . Panel (b): DF calculation at $U=12t$ and $T=t/6$, using a fit of $t=3$ meV.}
\label{Fig:compare}
\end{figure}

Figure~\ref{Fig:spin}(d), (e), and (f) show the imaginary part of the dynamical magnetic susceptibility $\mathrm{Im}\chi_m(\bq,\omega)$ as a function of energy $\omega$ and momentum ${\bf q}$ along a high symmetry path in the Brillouin zone (see labels in panel (a)). White dots show the energy $\omega_m({\bf q})$ of the maximum intensity at each momentum, referred to as the spin-wave dispersion. 
It is clear that, except for momenta near the $\mathrm{\Gamma}$ point, the intensity of spin excitations is enhanced and the spin excitation energy decreases as $U$ increases. At $U=6t$ there is no dominant spin excitation and spin fluctuations occur in a large part of the Brillouin zone. At $U=8.2t$ and $U=12t$ the spin excitation energy at the K point is smaller compared to other momenta, and most spin fluctuations occur at the K point. For $U=8.2t$ and $U=12t$, the spin excitation energy at the $\mathrm{\Gamma}$ point is nonzero, violating the total spin conservation. 
This is an artifact of the DF approximation \cite{LeBlanc19,HafermannPRB2014}. 

Figure~\ref{Fig:trace_k}(a) shows the spin excitation energy $\omega_m({\bf q})$ at the K point as a function of $U$. $\omega_m(\mathrm{K})$ approaches to two different values at small and large $U$, and a sharp decrease occurs as $U$ increases from $6t$ to $7t$. Figure~\ref{Fig:trace_k}(b) shows $\mathrm{Im}\chi_m({\bf q},\omega)$ at ${\bf q}=\mathrm{K}$ for various different $U$ values. The maximum value of $\mathrm{Im}\chi_m(\mathrm{K},\omega)$ increases very little as $U$ increases from $6t$ to $7t$, while it increases rapidly as $U$ continues to increase. These results suggest that spin fluctuations start to condense at the K point around $U=7t$. The \textcolor{black}{density of states} (DOS) at the Fermi surface, plotted in Fig.~\ref{Fig:trace_k}(a), shows that the system is still metallic at $U=7t$. Our results therefore suggest that the strong spin fluctuations not only exist in the insulating phase but also extend into the metallic phase.


In Fig.~\ref{Fig:compare} we compare our numerical data with the experimental magnetic susceptibilities obtained from $\mathrm{Ba}_8\mathrm{CoNb}_6\mathrm{O}_{24}$\cite{Rawl17}. To the left of the green line, no experimental data is available. Our simulations were obtained for $U=12t$ and $T=t/6$, and we set $t=3$ meV to fit the experimental data. Both our numerical data and the experimental data show that the intensity of the spin excitation around the K point is strong. We also note that the spin gap at the K point and the spin excitation energy at the M point are similar. In Ref.~\cite{Rawl17}, a Heisenberg model was used to fit Fig.~\ref{Fig:compare}(a), giving an estimated spin-spin interaction $J=0.144$ meV. In our case $J=1$ meV if $J=4t^2/U$ is used. Two sources contribute to this discrepancy. First, an energy scaling factor inherent to the DF approximation \cite{LeBlanc19} may change the overall energy scale of the DF results, leading to a larger fit value. More importantly, the spin spectra of the Hubbard and Heisenberg models, when compared at zero temperature on a $3\times3$ lattice using ED, show agreement only for $U>20t$ and differ markedly at $U=12t$ \citep{Supplement}.

\begin{figure}[t]
\center{\includegraphics[width=0.9\columnwidth]{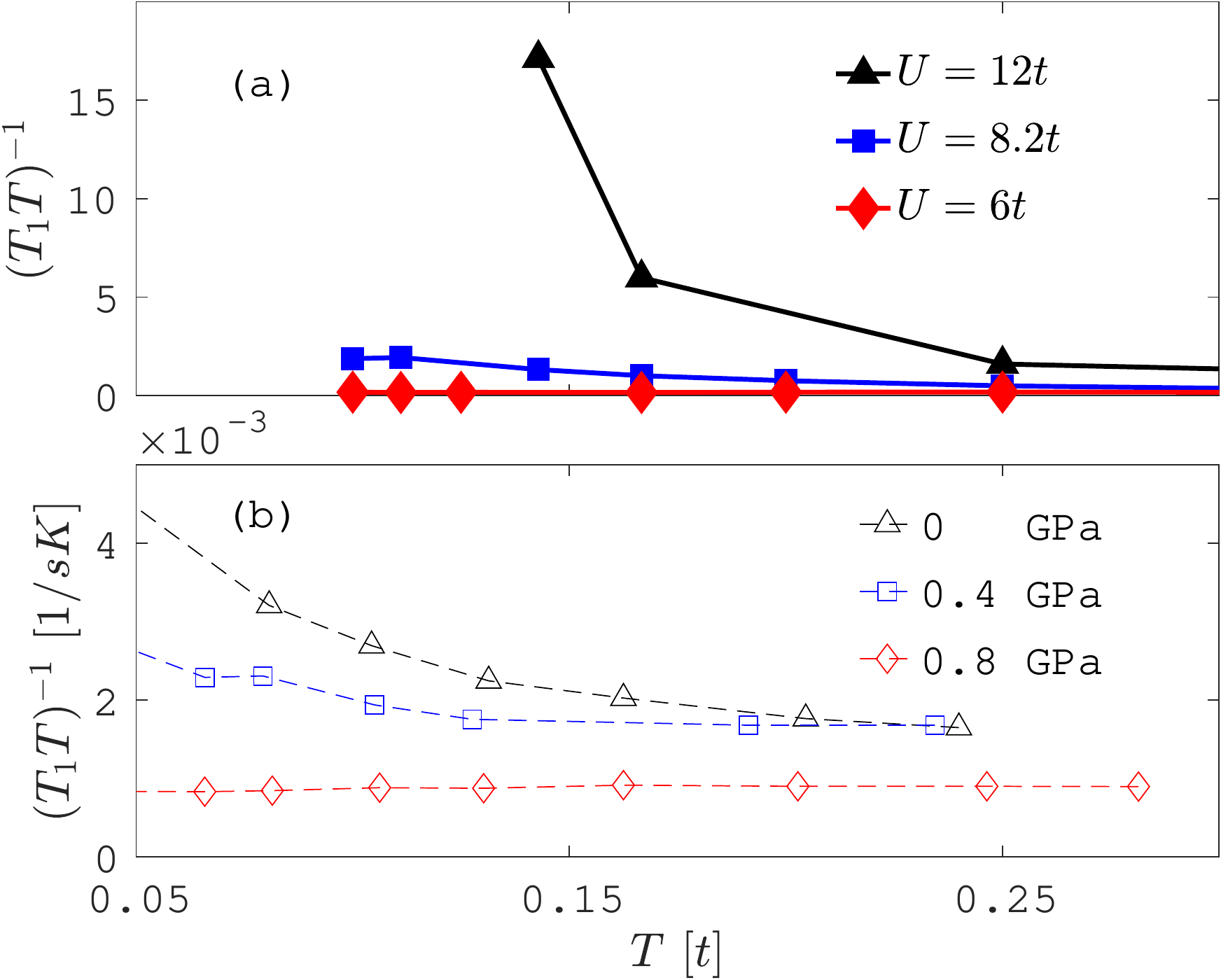}}
\caption{Panel (a) Spin-lattice decay rate \textcolor{black}{$(T_1T)^{-1}$} as a function of temperature $T/t$ for different $U$ values. Panel (b) shows \textcolor{black}{$(T_1T)^{-1}$} from $^{1}$H NMR measurements of \textcolor{black}{$\kappa$-(ET)$_2$Cu$_2$(CN)$_3$} \cite{Kurosaki05}. \textcolor{black}{The hopping integral $t$ for \textcolor{black}{$\kappa$-(ET)$_2$Cu$_2$(CN)$_3$} is about 0.055 eV} \cite{doi:10.1143/JPSJ.78.083710}.}
\label{Fig:T1}
\end{figure}

\begin{figure*}[t]
\center{\includegraphics[width=\textwidth]{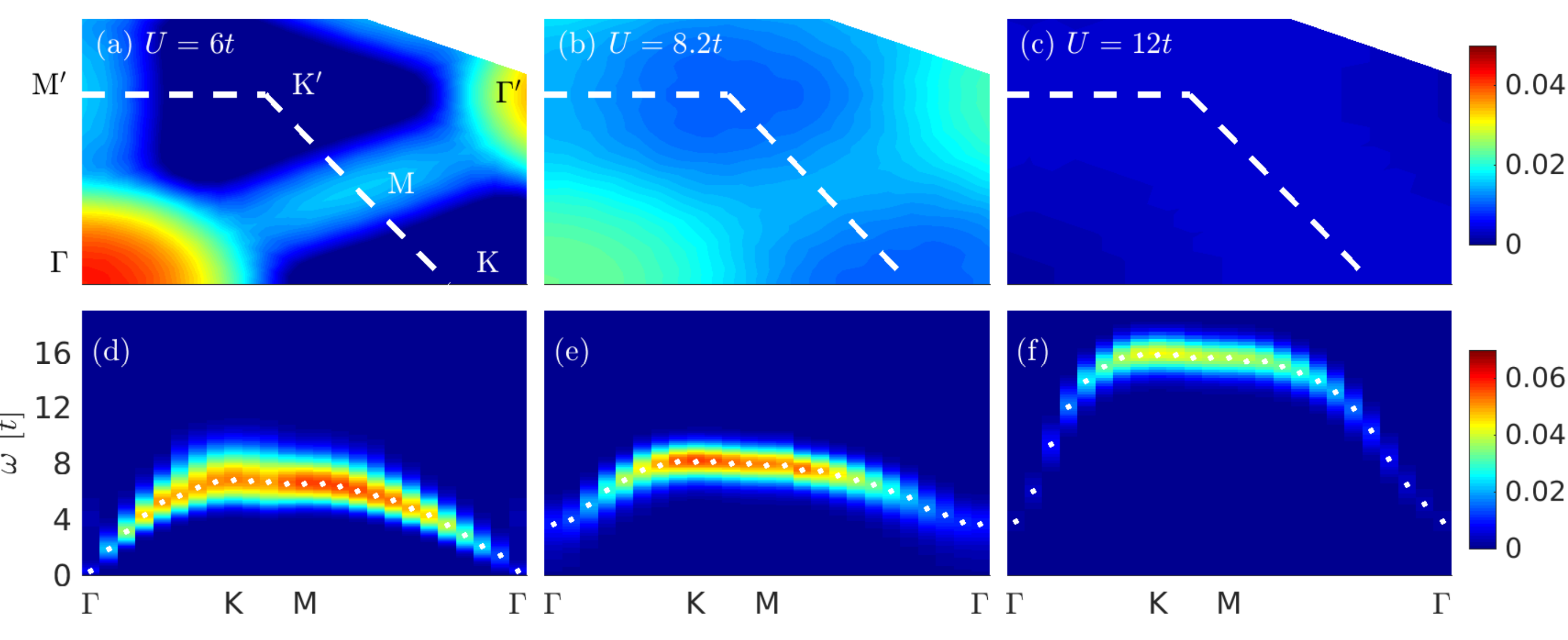}}
\caption{Panels (a), (b), and (c): Static charge susceptibility in momentum space, $\chi_c({\bf q},\mathrm{i}\nu_0)$, for $U=6t$, $8.2t$, and $12t$ at $T=t/6$. The white dashed line delineates the Brillouin zone boundary. Panels (d), (e), and (f): Imaginary part of the dynamical charge susceptibility $\mathrm{Im}\chi_c({\bf q},\omega)$ along the high symmetry direction for the same three values of $U$. White dots: peak position of the spectra.
\label{Fig:charge} }
\end{figure*}

Figure~\ref{Fig:T1}(a) plots the spin-lattice relaxation rate \textcolor{black}{$(T_1T)^{-1}$} as a function of temperature for the three values of $U$. \textcolor{black}{$(T_1T)^{-1}$} is calculated via $\underset{\omega \rightarrow 0}{\lim}\sum_{\bq} \frac{\mathrm{Im}\chi_m({\bq},\omega)}{\omega}$ \cite{Chen2017}. 
It is shown that \textcolor{black}{$(T_1T)^{-1}$} is enhanced as $U$ increases.
At $U=12t$, \textcolor{black}{$(T_1T)^{-1}$} increases rapidly as temperature decreases, indicating the formation of a magnetic order at low temperature with the transition temperature $T\approx 0.125t$. For $U=8.2t$, \textcolor{black}{$(T_1T)^{-1}$} increases very \textcolor{black}{slowly} at low temperature, consistent with the previous result that the magnetic order is absent at low temperature \citep{Li14}. The increase of \textcolor{black}{$(T_1T)^{-1}$} also implies that spin fluctuations are not negligible.
At $U=6t$, \textcolor{black}{$(T_1T)^{-1}$} is almost independent of temperature, consistent with the weak spin fluctuations shown in Fig.~\ref{Fig:spin}(d).

Figure~\ref{Fig:T1}(b) shows the spin-lattice relaxation rate of \textcolor{black}{$\kappa$-(ET)$_2$Cu$_2$(CN)$_3$} measured under different pressures, extracted from Ref.~\cite{Kurosaki05}. 
The $x$-axis has been rescaled by $t$, which is about 0.055 eV for \textcolor{black}{$\kappa$-(ET)$_2$Cu$_2$(CN)$_3$} \cite{doi:10.1143/JPSJ.78.083710}. At 0 \textcolor{black}{GPa}, corresponding to the spin liquid region, \textcolor{black}{$(T_1T)^{-1}$} monotonously increases as temperature decreases.
At 0.4 \textcolor{black}{GPa}, corresponding to the metallic phase near the phase boundary, \textcolor{black}{$(T_1T)^{-1}$} increases as $T$ decreases and reaches the maximum value at about $0.03t$. Continuing to decrease temperature, \textcolor{black}{$(T_1T)^{-1}$} decreases due to the appearance of a superconducting state, which is not plotted in Fig.~\ref{Fig:T1}(b) and absent in our calculations. At 0.8 \textcolor{black}{GPa}, \textcolor{black}{$(T_1T)^{-1}$} is independent of temperature. We notice that the temperature dependent behavior of \textcolor{black}{$(T_1T)^{-1}$} for these three pressures is similar to that for the three values of $U$ we calculated. The main difference is the behavior of \textcolor{black}{$(T_1T)^{-1}$} at high temperature. In experiment, \textcolor{black}{$(T_1T)^{-1}$} has a smaller value at 0.8 \textcolor{black}{GPa} than that for 0 \textcolor{black}{GPa} and 0.4 \textcolor{black}{GPa} at high temperature. In our calculations, \textcolor{black}{$(T_1T)^{-1}$} approaches the same value at high temperature. 
This difference may be due to pressure changes of the lattice geometry in \textcolor{black}{$\kappa$-(ET)$_2$Cu$_2$(CN)$_3$} \cite{clay2019}.

We next examine the charge properties. Figure~\ref{Fig:charge}(a), (b), and (c) show the static charge susceptibility $\chi_c({\bf q},\mathrm{i}\nu_0)$ for the same three values of $U$. It is clearly seen that $\chi_c({\bf q},\mathrm{i}\nu_0)$ is suppressed as $U$ increases and is invisible in the insulator ($U=12t$). 
At $U=6t$ the maximum value of $\chi_c({\bf q},\mathrm{i}\nu_0)$ is located at the $\Gamma$ point, indicating a uniform charge distribution. $\chi_c({\bf q},\mathrm{i}\nu_0)$ along the $\Gamma^{\phantom\prime} \rightarrow \Gamma^\prime$ direction ($\Gamma^\prime$ is the $\Gamma$ point in the second Brillouin zone) is larger compared to the other momenta. These features are weaker at $U=8.2$ and invisible at $U=12t$.

Figure~\ref{Fig:charge}(d), (e), and (f) plot the imaginary part of the charge susceptibility $\mathrm{Im}\chi_c({\bf q},\omega)$ for these three values of $U$. 
At $U=6t$ there is no charge gap at the $\Gamma$ point and the maximum energy of the charge excitation is located around the Brillouin zone boundary, corresponding to the charge excitation from the bottom to the top of the band. 
Little change is visible near the crossover, $U=8.2t$.  The nonzero charge excitation at the $\Gamma$ point is because DF violates the total charge conservation \cite{LeBlanc19,HafermannPRB2014}. 
We also note that the maximum energy of the charge excitation does not change much as $U$ increases before the Mott gap is opened, while it increases rapidly as the gap is opened. Our predicted charge spectra may be observed in momentum-resolved electron-loss spectroscopy measurements on triangular compounds.

Finally, we compare our magnetic and charge susceptibilities to the bare susceptibility $\mathrm{Im}\chi_0({\bf q},\omega)$ \cite{Supplement}, which is evaluated by a multiplication of two Green's functions \citep{Supplement}. The low energy spectra of $\mathrm{Im}\chi_0({\bf q},\omega)$ and $\mathrm{Im}\chi_m({\bf q},\omega)$ are consistent at $U=6t$ but inconsistent at $U=8.2$ and $U=12t$. The high energy spectra of $\mathrm{Im}\chi_0({\bf q},\omega)$ are consistent with $\mathrm{Im}\chi_c({\bf q},\omega)$ only near the Brillouin zone boundary for these three values of $U$. These discrepancies suggest that the many-body effects or vertex corrections are essential.

{\it Summary}. We have studied the  Hubbard model on a triangular lattice and presented the momentum and energy dependence of the spin and charge spectra in the metallic, Mott insulating, and crossover regimes. We find that the strong spin fluctuations at the K point at low energy exist in not only the insulator but also the metallic phase. We also compared our simulated data of neutron spectroscopy and relaxation times to inelastic neutron spectroscopy experiments on the triangular material $\mathrm{Ba}_8\mathrm{CoNb}_6\mathrm{O}_{24}$ and to the relaxation times on \textcolor{black}{$\kappa$-(ET)$_2$Cu$_2$(CN)$_3$}.
 
Our work employed the fermion Hubbard model instead of spin models which are typically used to study spin excitation spectra in frustrated systems. Unlike spin models, which are a low energy limit of the Hubbard model at large $U$ ($U>20t$), our results are valid both in the metallic and the insulating regime.

\begin{acknowledgments}
This work was supported by the National Science Foundation (NSF)
under Grant No. DMR-1606348. This work used resources of the Extreme Science and Engineering Discovering Enviroment (XSEDE) under Grant No. TG-DMR130036.
\end{acknowledgments}
\bibliographystyle{apsrev4-1}
\bibliography{main}

\newpage\phantom{blabla}
\pagebreak
\onecolumngrid
\begin{center}
{\large Supplementary material for ``Magnetic and charge susceptibilities in the half-filled triangular lattice Hubbard model"}

Shaozhi Li$^1$ and Emanuel Gull$^{1,2}$

$^1${\it Department of Physics, University of Michigan,  Ann Arbor, Michigan 48109, USA}\\
$^{1,2}${\it Center for Computational Quantum Physics, The Flatiron Institute, New York, New York, 10010, USA}
\end{center}

\newcommand{\beginsupplement}{%
	\setcounter{table}{0}
	\renewcommand{\thetable}{S\arabic{table}}%
	\setcounter{figure}{0}
	\renewcommand{\thefigure}{S\arabic{figure}}%
	\setcounter{equation}{0}
	\renewcommand{\theequation}{S\arabic{equation}}%
	\setcounter{section}{0}
	\renewcommand{\thesection}{S\arabic{section}}%
	\renewcommand{\thesubsection}{S\arabic{section}.\arabic{subsection}}%
}
\renewcommand*{\citenumfont}[1]{S#1}
\renewcommand*{\bibnumfmt}[1]{[S#1]}
\titlespacing*{\subsection} {0pt}{25pt}{0pt}

\beginsupplement
\section{Metal-Insulator crossover}
\begin{figure}[ht]
\center{\includegraphics[width=0.5\columnwidth]{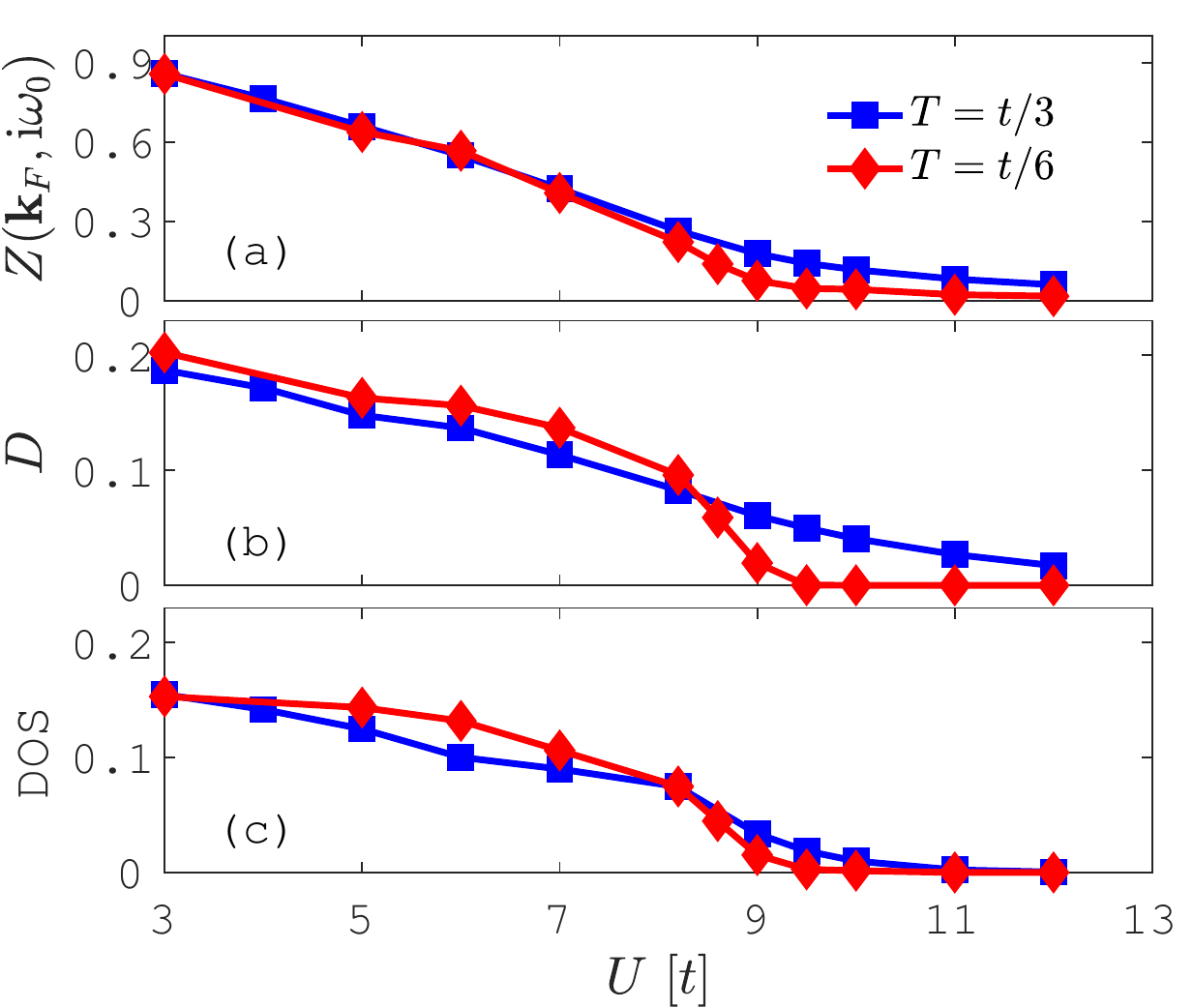}}
\caption{Metal-Insulator crossover as a function of $U$, for temperature $T=t/3$ and $T=t/6$. (a) evolution of the quasiparticle weight estimate $Z$ at momentum ${\bf k}_F=(5\pi/6,0)$. (b) double occupancy. (c) local density of states.
\label{Fig:InteractionTuning}}
\end{figure}

Fig.~\ref{Fig:InteractionTuning} shows the quasiparticle weight $Z$, the double occupancy $D=\langle \hat{n}_{\uparrow}\hat{n}_{\downarrow}\rangle$, and the local density of states as a function of interaction strength at temperature $T=t/3$ and $T=t/6$. Both of these temperatures are in a crossover regime above the metal-insulator phase transition. 
$Z$ is approximately determined as $Z({\bf k}_F,\mathrm{i}\omega_0)=1/(1-\mathrm{Im}\Sigma({\bf k}_F,\mathrm{i}\omega_0)/\omega_0)$,
where $\Sigma({\bf k}_F,i\omega_0)$ is the self-energy at the lowest Matsubara frequency $\omega_0$, and ${\bf k}_F=(5\pi/6a,0)$ is a momentum on the Fermi surface of the non-interacting system.
The double occupancy is obtained via $D=\frac{1}{2}\left[\chi_c^{\mathrm{loc}}(\tau=0)-2\chi_s^{\mathrm{loc}}(\tau=0) + 2\langle \hat{n}_{\uparrow}\rangle \langle \hat{n}_{\downarrow}\rangle\right]$, where $\chi_{c(s)}^{\mathrm{loc}}(\tau)$ is the local charge (spin) susceptibility.
Figure~\ref{Fig:InteractionTuning}(c) plots the local density of states (DOS) at the Fermi surface, obtained via analytic continuation of the local electron Green's function.
All three quantities are large at small $U$, consistent with metallic behavior, and approach a small value at a large $U$, consistent with insulating behavior.

\section{Spectral function}
\begin{figure}[h]
\center{\includegraphics[width=\textwidth]{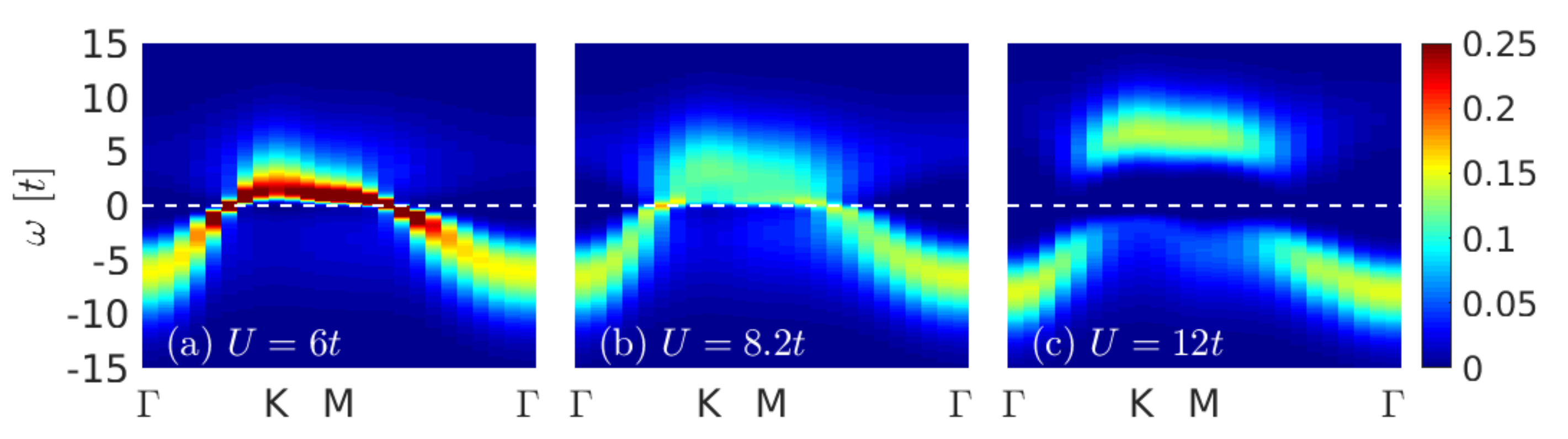}}
\caption{Momentum dependence of the spectral function $A({\bf k},\omega)$ for (a) $U=6t$, (b) $U=8.2t$, and (c) $U=12t$ at $T=t/6$. The white dashed line shows the Fermi surface.
\label{Fig:spectral function} }
\end{figure}

Figure~\ref{Fig:spectral function}(a)-(c) show the momentum resolved single-particle spectral functions $A({\bf k},\omega)$ for $U=6t$, $8.2t$, and $12t$ at $T=t/6$. The white dashed line indicates the Fermi surface. At $U=6t$, when it is a metal, there is a strong quasiparticle peak at the Fermi surface. At $U=8.2t$, the intensity of this quasiparticle peak is suppressed and the lower (upper) Hubbard band forms around K ($\Gamma$) point. At $U=12t$, a Mott gap is fully opened. Our calculations show that the Mott gap is opened as $U>10t$ [see Fig.~\ref{Fig:InteractionTuning}]. We notice that there is no superstructure in Fig.~\ref{Fig:spectral function} (c) as long-ranged magnetic order is absent. All these features we observed in the spectral function are consistent with previous cluster perturbation theory results \cite{Misumi17}.

\section{Charge-density-wave and magnetic susceptibilities}
The charge-density-wave (CDW) susceptibility is defined as
\begin{eqnarray}
\chi_c({\bf q},\mathrm{i}\nu_n)=\int_0^{\beta}d\tau e^{\mathrm{i\nu_n\tau}} \left[\langle \rho_{\bf q}(\tau) \rho_{-{\bf q}}(0) \rangle - \langle \rho_{\bf q}(\tau)  \rangle \langle \rho_{-{\bf q}}(0) \rangle \right],
\end{eqnarray}
where $\rho_{\bf q}(\tau)=\sum_{{\bf r},\sigma} e^{\mathrm{i}{\bf q}\cdot{\bf r}}\hat{n}_{{\bf r},\sigma}(\tau)$, and $\hat{n}_{{\bf r},\sigma}^{\phantom\dagger}=c_{{\bf r},\sigma}^{\dagger}(\tau)c_{{\bf r},\sigma}^{\phantom\dagger}(\tau)$. 

The magnetic susceptibility is defined as
\begin{eqnarray}
\chi_m({\bf q},\mathrm{i}\nu_n)=\int_0^{\beta}d\tau e^{\mathrm{i\nu_n\tau}}\langle S^z_{\bf q}(\tau) S^z_{-{\bf q}}(0) \rangle,
\end{eqnarray}
where $S^z_{\bf q}(\tau)=\frac{1}{2}\sum_{{\bf r}} e^{\mathrm{i}{\bf q}\cdot{\bf r}}\left[\hat{n}_{{\bf r},\uparrow}(\tau)-\hat{n}_{{\bf r},\downarrow}(\tau)\right]$.

\section{Comparison between the Hubbard and Heisenberg models}
\begin{figure}[t]
\center{\includegraphics[width=\textwidth]{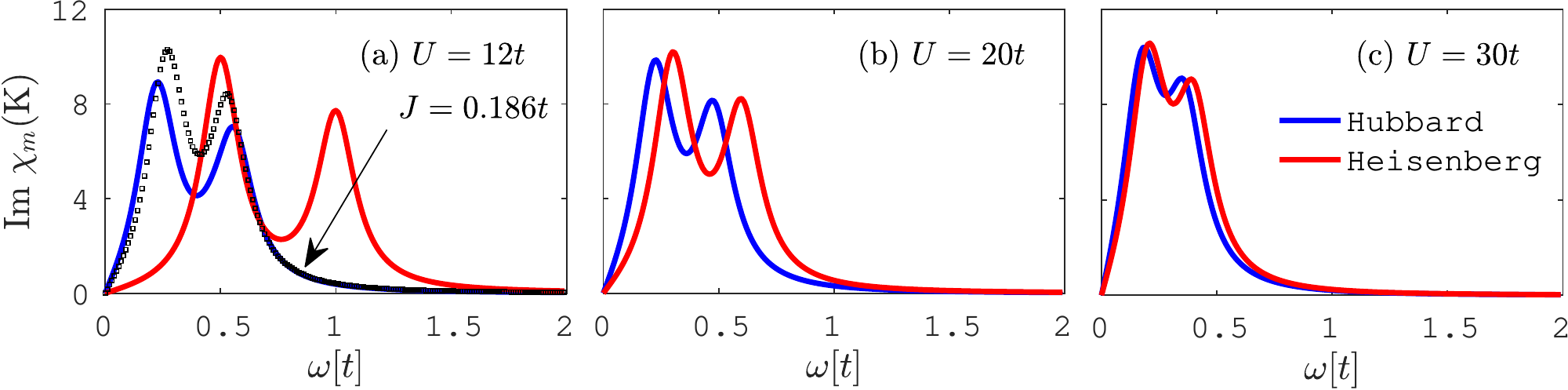}}
\caption{ The imaginary part of spin susceptibilities $\mathrm{Im}\chi_m({K},\omega)$ at zero temperature for (a) $U=12t$, (b) $U=20t$, and (c) $U=30t$. Blue lines represent the Hubbard model, and red lines represent the Heisenberg model with $J=4t^2/U$. Black dots in pannel (a) show results of the Heisenberg model with $J=0.186t$.
\label{Fig:comparison} }
\end{figure}
In the strong Coulomb interaction limit ($U\gg t $), the Hubbard Hamiltonian simplifies to the Heisenberg Hamiltonian with a spin-spin interaction $J=4t^2/U$. To valid this simplification, we calculate the imaginary part of spin susceptibilities $\mathrm{Im}\chi_m({\bf q},\omega)$ on a $3\times 3$ lattice using exact diagonalization at zero temperature for the Hubbard and Heisenberg models, respectively. Figure~\ref{Fig:comparison} plot $\mathrm{Im}\chi_m({K},\omega)$ as a function of frequency $\omega$ at $U=12t$, $20t$, and $30t$. At $U=12t$ ($J=0.333t$) $\mathrm{Im}\chi_m({K},\omega)$ of the Hubbard model is located at lower energy compared to the Heisenberg model, and can be fitted by the Heisenberg model with $J=0.186t$, implying that a spin-spin interaction of $4t^2/U$ is overestimated by a factor of two. $\mathrm{Im}\chi_m({K},\omega)$ of the Heisenberg model moves toward to $\mathrm{Im}\chi_m({K},\omega)$ of the Hubbard model as $U$ increases, and they overlap as $U>30t$. These results indicate that the low energy physics of the Hubbard model can be described by the Heisenberg model only as $U>20t$.

\section{Bare susceptibility}
\begin{figure}[!ht]
\center{\includegraphics[width=\textwidth]{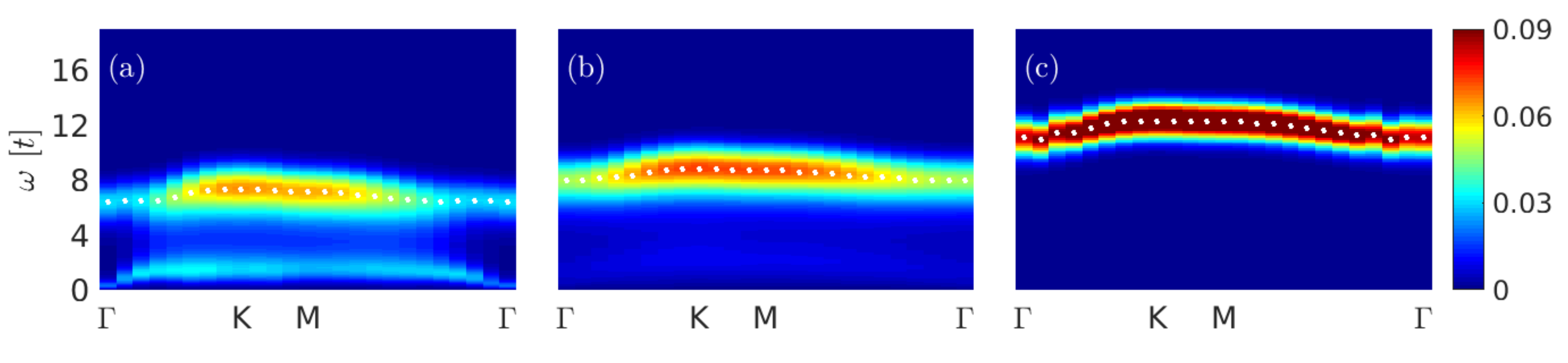}}
\caption{Momentum dependence of the imaginary part of the bubble susceptibility $\chi_0({\bf q},\omega)$ for (a) $U=6t$, (b) $U=8.2t$, and (c) $U=12t$ at $T=t/6$. White dots: peak position of the spectra.
\label{Fig:bubble} }
\end{figure}

The bare susceptibility is defined as 
\begin{eqnarray}
\chi_0({\bf q},\mathrm{i}\nu_n)=-\frac{1}{N\beta}\sum_{{\bf k},\omega_m} G_{\sigma}({\bf k},\mathrm{i}\omega_m)G_{\sigma}({\bf k}+{\bf q},\mathrm{i}\omega_m+\mathrm{i}\nu_n),
\end{eqnarray}
where $\omega_m$ ($\nu_n$) is the fermion (bosonic) Matsubara frequency. We use the maximum-entropy method to perform the analytical continuation. Figure~\ref{Fig:bubble} shows the imaginary part of the bubble susceptibility $\chi_0({\bf q},\omega)$. At $U=6t$ the spectrum is split into two branches. The low-energy spectrum corresponds to the spin excitation as shown in Fig.~1(b) in the main text; the high-energy spectrum corresponds to the charge excitation as shown in Fig.~4(b) in the main text. The intensity of the low-energy spectrum is suppressed at $U=8.2t$. At $U=12t$ the low-energy spectrum completely disappears. The high-energy spectrum has a weak momentum dependence and moves to higher energy as $U$ increases. All these results are different from magnetic and charge susceptibilities shown in the main text, implying that vertex corrections are important.


\end{document}